\documentclass{article}
\usepackage{latexsym}
\usepackage{graphicx}

\begin{document}

\begin{center}
\textbf{Colossal Nernst power factor in topological semimetal 
NbSb$_{2}$}
\end{center}

Peng Li$^{1,2}$, Pengfei Qiu$^{1,2}$*, Qing Xu$^{3}$, Jun Luo$^{1,2}$, Yifei 
Xiong$^{1,2}$, Jie Xiao$^{1}$, Niraj Aryal$^{4}$, Qiang Li$^{4,5}$, Lidong 
Chen$^{1,2}$, and Xun Shi$^{1,2}$*\\

$^{1}$State Key Laboratory of High Performance Ceramics and Superfine 
Microstructure, Shanghai Institute of Ceramics, Chinese Academy of Science, 
Shanghai 200050, China

$^{2}$Center of Materials Science and Optoelectronics Engineering, 
University of Chinese Academy of Sciences, Beijing 100049, China

$^{3}$Key Laboratory of Infrared Imaging Materials and Devices, Shanghai 
Institute of Technical Physics, Chinese Academy of Sciences, Shanghai 
200083, China

$^{4}$Condensed Matter Physics and Materials Science Division, Brookhaven 
National Laboratory, Upton, New York 11973-5000, USA

$^{5}$Department of Physics and Astronomy, Stony Brook University, Stony 
Brook, New York 11794-3800, USA\\

\textit{*E-mail: qiupf@mail.sic.ac.cn; xshi@mail.sic.ac.cn}\\

\textbf{Abstract}
\newline

Today solid-state cooling technologies below liquid nitrogen boiling 
temperature (77 K), crucial to quantum information technology and probing 
quantum state of matter, are greatly limited due to the lack of good 
thermoelectric and/or thermomagnetic materials. Here, we report the 
discovery of colossal Nernst power factor of 3800\texttimes 10$^{-4}$ W 
m$^{-1}$ K$^{-2}$ under 5 T at 25 K and high Nernst figure-of-merit of 
71\texttimes 10$^{-4}$ K$^{-1}$ under 5 T at 20 K in topological semimetal 
NbSb$_{2}$ single crystals. The observed high thermomagnetic performance is 
attributed to large Nernst thermopower and longitudinal electrical 
conductivity, and relatively low transverse thermal conductivity. The large 
and unsaturated Nernst thermopower is the result of the combination of 
highly desirable electronic structures of NbSb$_{2}$ having compensated high 
mobility electrons and holes near Fermi level and strong phonon-drag effect. 
This discovery opens an avenue for exploring material option for the 
solid-state heat pumping below liquid nitrogen temperature.

\newpage

\section*{Introduction}

Capable of converting heat into electricity and\textit{ vice versa} without moving parts and 
green house emission, thermoelectricity plays an important role in solid 
state energy harvesting and cooling$^{1-4}$. Current thermoelectric (TE) 
technologies are largely developed for applications around room temperature 
(cooling/heating) and above (e.g. waste heat recovery), primarily due to the 
state-of-the-art TE materials exhibiting large electronic entropy, or a 
large TE power factor - a measurement of entropy transfer capability$^{5}$. 
However, there is a demand now for TE applications at low temperatures, 
especially near or below liquid helium boiling point (4.2 K) heightened by 
applications in exploring quantum state of matters$^{6}$, quantum 
information science and technologies$^{7}$, and space science and 
technologies$^{8}$, among others.

TE refrigeration used today is based on the Peltier effect that has the 
advantages of accurate and fast temperature control, and nearly 
maintenance-free. As shown in Fig. 1a, when a longitudinal current flows 
through a thermopile, a longitudinal temperature gradient is formed, 
yielding the reduction of temperature at one end of the thermopile. Vast 
majority of TE devices used today have the longitudinal configurations (Fig. 
1a) made of n- and p-type elements connected in series where electrical and 
thermal resistive contact are primary sources of reduced 
efficiency$^{9-12}$. A maximum temperature drop of around 70 K has been 
achieved in typical Bi$_{2}$Te$_{3}$-based TE devices at room temperature by 
the Peltier effect$^{4}$. However, such technology is greatly limited at 
temperatures below liquid nitrogen boiling point (\textasciitilde 77 K) due 
to the lack of high-performance TE materials in the low temperature range 
(Fig. 1b and 1c). The physical reason behind this phenomenon is understood 
as follows. It has been argued that a good conventional TE semiconductor 
usually has a band gap \textasciitilde 10$ k_{B}T$ (where$ k_{B}$ is the 
Boltzmann constant and$ T$ is the working temperature)$^{13}$. In order to have 
it efficiently work below 77 K, the band gap should be less 66 meV. This 
leads to the great difficulty of finding the potential high-performance TE 
materials for Peltier refrigeration in the low temperature range since the 
semiconductors with band gap less than 66 meV are rare.

Ettingshausen refrigerator is a transverse TE device that provides cooling 
orthogonal to the applied voltage that greatly simplifies a thermopile 
structure, in which only the electrical contact is required at colder side 
of the thermoelectric material that does not require compatible p- and 
n-type elements, and can reduce the thermal resistance. The Ettingshausen 
effect is shown in Fig. 1a. When a longitudinal current (along the $y$- 
direction) flows through a thermomagnetic material under magnetic field, a 
transverse temperature gradient (along the $x$- direction) will be formed, 
yielding the reduction of temperature at the material's transverse side 
surface. The cooling efficiency of Ettingshausen refrigeration is determined 
by material's electrical and thermal transport properties in orthogonal 
directions, which can be evaluated by a comprehensive parameter named as the 
Nernst figure-of-merit$^{14}$, 
$z_{\mathrm{N}}\mathrm{=}\frac{S_{yx}^{\mathrm{2}}\sigma_{yy}}{\kappa 
_{xx}}$, where$ S_{yx}$ is the Nernst thermopower, $\sigma_{yy}$ is the 
longitudinal electrical conductivity, and $\kappa_{xx}$ is the transverse 
thermal conductivity, and magnetic field is along the $z$-direction, 
respectively. In contrast to the longitudinal TE power factor \textit{PF} 
($=S^{2}\sigma $, where $S$ and $\sigma $are the Seebeck thermopower and 
electrical conductivity along the same direction, respectively), the Nernst 
power factor ${PF}_{\mathrm{N}}=S_{yx}^{2}\sigma_{yy}$ is used to determine 
the transverse pumping power. Electrons and holes moving in the opposite 
direction driven by the longitudinal current, can carry both charge and 
energy in the same transverse direction synergistically under magnetic 
field, resulting in doubling of transverse temperature gradient. Therefore, 
semimetals with zero band gap or slightly band overlap are particularly 
suitable for the Ettingshausen cooling at low temperatures below 77 K.

Although the Ettingshausen effect was discovered in 1886$^{15}$, the 
Ettingshausen refrigeration has progressed far less than Peltier 
refrigeration. For a long time, the investigation is only limited in a few 
thermomagnetic materials, such as Bi-Sb alloys$^{16,\, 17}$ and In-Sb 
alloys$^{18}$. The peak $z_{N}$ values of single-crystalline 
Bi$_{97}$Sb$_{3}^{16}$ and Bi$_{99}$Sb$_{1}^{17}$ are 55\texttimes 
10$^{-4}$ K$^{-1}$ under 1 T and 29\texttimes 10$^{-4}$ K$^{-1}$ under 0.75 T 
at 115 K, respectively (Fig. 1b). Recently, the discovery of topological 
semimetals with high carrier mobility has rejuvenated the investigation of 
Ettingshausen effect$^{19-27}$. It is noted that the Dirac-like linear 
electronic band dispersion near Fermi level in topological semimetals$^{28}$ 
can lead to an energy-independent electronic density of states that 
increases linearly with magnetic field, thus create huge electronic 
entropy$^{20,\, 29}$. Indeed, the peak $z_{N}$ of Dirac semimetal ZrTe$_{5}$ 
was reported to reach 10.5\texttimes 10$^{-4}$ K$^{-1}$ under 13 T at 120 
K$^{21}$. Nodal-line semimetal PtSn$_{4}$ has the peak $z_{N}$ of 8\texttimes 
10$^{-4}$ K$^{-1}$ under 9 T at 10 K$^{22}$. Most recently, Pan et al. 
reported an ultrahigh$ z_{N}$ of 265\texttimes 10$^{-4}$ K$^{-1}$ under 9 T at 
11.3 K in single-crystalline Weyl semimetal WTe$_{2}^{27}$. This value is 
already much higher than that of Bi-Sb alloys (Fig. 1b), which was recently 
shown to be also a topological semimetal in specific chemical composition 
range after all$^{30}$. These results motivate the discovery of new 
thermomagnetic materials with high $z_{N}$ below liquid nitrogen temperature 
from topological semimetals.

In this work, we report that topological semimetal NbSb$_{2}$ single crystal 
is a promising high-performance thermomagnetic material with a colossal 
\textit{PF}$_{N}$ of 3800\texttimes 10$^{-4}$ W m$^{-1}$ K$^{-2}$ under 5 T at 25 K 
(Fig.1c) and a high $z_{N}$ of 71\texttimes 10$^{-4}$ K$^{-1\, }$under 5 T at 
20 K (Fig. 1b), much higher than most TE and thermomagnetic materials below 
77 K. We found that the performance in NbSb$_{2}$ benefits from the 
combination of nearly identical electron and hole concentrations, high 
electron/hole carrier mobilities, and additional phonon-drag effect. \\

\section*{Results}

\textbf{Crystal structure.} NbSb$_{2}$ is a topological semimetal$^{31}$. It 
crystalizes in centrosymmetric monoclinic structure with the space group of 
$C_{2/m}$. The schematics of its crystal structure is shown in Fig. 2a. The 
Nb atom is enclosed in a hendecahedron composed of Sb atoms. The 
hendecahedrons are connected with each other in the way of face to face 
along the $b$-axis and edge to edge along the $c$-axis, forming an atomic layer 
parallel to \textit{bc} plane. The lattice parameters for NbSb$_{2}$ are $a \quad =$10.239 
{\AA}, $b \quad =$3.632 {\AA}, $c =$ 8.333 {\AA}, and $\beta \quad =$ 120.07\textdegree 
$^{32}$. Fig. 2b shows the NbSb$_{2\, }$single crystal grown by the chemical 
vapor transport method. The NbSb$_{2\, }$single crystal has a bar-like shape 
with the length about 7 mm and the width about 1-2 mm. The X-ray 
characterization performed on the upper surface (Supplementary Fig. 1a) 
shows that strong (200), (400), and (600) diffraction peaks are observed, 
indicating the high quality of our NbSb$_{2\, }$single crystal. 
Supplementary Fig. 1b shows that Nb and Sb are homogeneously distributed 
inside the matrix, consistent with the pure phase detected by XRD 
measurement.

\textbf{Band structure.} Fig. 2c shows the calculated band structure of 
NbSb$_{2}$ with the inclusion of spin-orbit coupling (SOC) effect. The Fermi 
level crosses the conduction band on the path from $L$ to $I$ and the valence band 
near $L$, rendering it a typical semimetal. The energy overlap between 
conduction band and valence band is about 350 meV. From the Fermi surface 
(FS) plotted in Fig. 2d, we can identify one electron pocket (blue shell) 
and one hole pocket (red shell) in the first Brillouin zone. The calculated 
FS area on the \textit{ab} plane is comparable to the experimentally measured area from 
the quantum oscillation measurement$^{33}$.$^{\, }$A$^{\, }$plot showing 
variation of the calculated FS area with chemical potential and comparison 
with the experimental value is shown in the Supplementary Fig. 2, with the 
details shown in Supplementary Note 1. The similarity between the calculated 
and measured FS areas provides validity to the density functional theory 
(DFT)-predicted electronic structure. The electron pocket and the hole 
pocket have nearly the same volume leading to well compensated electrons and 
holes near the Fermi level. Under orthogonal applied magnetic field and 
current, the electrons and holes in these pockets moving in the opposite 
direction along the longitudinal current are deflected in the same 
transverse direction, which can strengthen the Ettingshausen effect.

\textbf{Transport properties.} Supplementary Figs. 3a-b show the temperature 
dependences of adiabatic transverse electrical resistivity $\rho_{xx}$ 
and Hall resistivity $\rho_{yx}$ of single-crystalline NbSb$_{2}$ under 
different magnetic fields $B$. When $B =$ 0, the $\rho_{xx}$ rises with 
increasing temperature, showing typical metal-like conduction behavior. The 
$\rho_{xx}$ is \textasciitilde 2\texttimes 10$^{-9} \Omega $ m at 5 K, 
which is about 3-4 orders of magnitude lower than those of typical TE 
materials for Peltier refrigeration$^{4,\, 34}$. Upon applying magnetic 
field, the $\rho_{xx}$ at temperatures below 100 K is greatly increased, 
a characteristic feature of topological semimetals$^{28,\, 35}$. The 
magnetoresistance (MR) ratio of single-crystalline NbSb$_{2}$ under 9 T at 5 
K is 1.3\texttimes 10$^{5\, }${\%}, comparable to those of extremely large 
magnetoresistance (XMR) materials reported before, such as MR $=$ 
8.5\texttimes 10$^{5\, }${\%} for NbP under 9 T at 1.85 K$^{36}$, 
4.5\texttimes 10$^{5\, }${\%} for WTe$_{2}$ under 14.7 T at 4.5 K$^{35}$, and 
5\texttimes 10$^{5\, }${\%} for PtSn$_{4}$ under 14 T at 1.8 K$^{37}$. 
Supplementary Fig. 3b shows that the absolute value of Hall resistivity 
(\textbar $\rho_{yx}$\textbar ) firstly decreases with increasing 
temperature, reaches a minimum at about 100 K, and then increases at higher 
temperature. Under the same magnetic field, the \textbar $\rho 
_{yx}$\textbar is much lower than the $\rho_{xx}$.

To evaluate the Nernst figure-of-merit, we need to know the longitudinal 
conductivity $\sigma_{yy}$, which can be calculated by the equation
\begin{equation}
\label{eq1}
\sigma_{yy}\mathrm{=}\frac{\rho_{xx}}{\rho_{xx}\rho_{yy}-\rho_{yx}\rho 
_{xy}}=\frac{\rho_{xx}}{{\left( \rho_{yy} \mathord{\left/ {\vphantom {\rho 
_{yy} \rho_{xx}}} \right. \kern-\nulldelimiterspace} \rho_{xx} \right)\rho 
}_{xx}^{2}+\rho_{yx}^{2}}
\end{equation}
where $\rho_{yy}$ is the longitudinal electrical resistivity. The value 
of $\rho_{yy}$/$\rho_{xx}$ is determined by measuring the electrical 
resistivity along the $b$-axis ($\rho_{xx})$ and the electrical resistivity 
along the $c$-axis ($\rho_{yy})$ of a thin square single-crystalline 
NbSb$_{2}$ sample (Supplementary Figs. 4a-b). It seems that the electrical 
resistivities behavior of NbSb$_{2}$ is more anisotropic at low 
temperatures, but less anisotropic at room temperature. Under the assumption 
that -$\rho_{xy}$ is equal to $\rho_{yx}$, the $\sigma_{yy}$ under 
different magnetic fields is calculated and shown in Fig. 3a. The $\sigma 
_{yy}$ firstly increases with increasing temperature, reaches a maximum, 
and then decreases with further increasing the temperature. The temperature 
corresponding to the maximum $\sigma_{yy}$ is gradually shifted from 35 K 
under $B =$ 1 T to 85 K under $B =$ 9 T.

The adiabatic Seebeck thermopower $S_{xx}$ below 100 K is very small under 
$B =$ 0 T (Fig. 3b), with the absolute value \textbar $S_{xx}$\textbar less 
than 5 $\mu $V K$^{-1}$. Below 100 K, it increases modestly with increasing 
magnetic field, with the peak value around 20 $\mu $V K$^{-1}$ even under 
$B =$ 9 T. Above 100 K, the \textbar $S_{xx}$\textbar increases with 
increasing temperature, but the maximum is still much lower than those of 
conventional TE materials$^{34,\, 38-40}$. Such low $S_{xx}$ values are 
consistent with the semimetal feature of NbSb$_{2}$ (Fig. 2c).

Fig. 3c shows the temperature dependence of adiabatic Nernst thermopower 
$S_{yx}$ of single-crystalline NbSb$_{2}$. Under a magnetic field, the 
absolute value of $S_{yx}$ initially increases with increasing temperature, 
reaches the maximum value around 21 K, and then decreases at higher 
temperatures. Similar behavior is observed when the direction of magnetic 
field is reversed, with the sign of $S_{yx}$ is reversed accordingly. The 
maximum $S_{yx}$ is about 616 $\mu $V K$^{-1}$ under 9 T at 21 K, about 
thirty times of the maximum $S_{xx}$. Likewise, as shown in the Supplementary 
Note 2, the thermal Hall effect has little influence on the $S_{yx}$ 
measurement.

The adiabatic Nernst power factor \textit{PF}$_{N}$ ($=S_{yx}^{2}\sigma_{yy})$ of 
single-crystalline NbSb$_{2}$ under different magnetic fields is shown in 
Fig. 3d. The \textit{PF}$_{N}$ firstly increases with increasing temperature, reaches a 
peak around 25 K, and then decreases at higher temperatures. At $B \quad =$ 1 T, 
the \textit{PF}$_{N}$ reaches 1750\texttimes 10$^{-4}$ W m$^{-1}$ K$^{-2}$ at 25 K. As 
shown in Fig. 1c, this value is already much higher than the best Peltier\textit{ PF} of 
the TE materials, such as 41\texttimes 10$^{-4}$ W m$^{-1}$ K$^{-2}$ for 
Bi$_{2}$Te$_{3}^{34}$, 75\texttimes 10$^{-4}$ W m$^{-1}$ K$^{-2}$ for 
SnSe$^{41}$, and 25\texttimes 10$^{-4}$ W m$^{-1}$ K$^{-2}$ for 
Mg$_{3}$Sb$_{2}^{34}$. This result is very encouraging as many permanent 
magnets can easily provide 1 T magnetic field, thus utilizing 
single-crystalline NbSb$_{2}$ for the Ettingshausen cooling is practically 
viable. At $B =$ 5 T, the \textit{PF}$_{N}$ is further enhanced to 
3800\texttimes 10$^{-4}$ W m$^{-1}$ K$^{-2}$ at 25 K. As shown in$^{\, 
}$Fig. 1c, this value is much higher than those of single-crystalline 
PtSn$_{4}^{22}$ and single-crystalline Mg$_{2}$Pb$^{26}$. It is only lower 
than that for WTe$_{2}$, which has the \textit{PF}$_{N}$ up to 36000\texttimes 
10$^{-4}$ W m$^{-1}$ K$^{-2}$ under 9 T at 15.9 K$^{27}$. 

Fig. 3e shows the adiabatic transverse thermal conductivity $\kappa_{xx}$ 
of single-crystalline NbSb$_{2}$ from 5 to 300 K measured by using the 
four-probe method. At $B =$ 0, the $\kappa_{xx} $increases with increasing 
temperature, reaches a peak of 90 W m$^{-1}$ K$^{-1}$ around 30 K, and then 
decreases with further increasing temperature. At 300 K, the $\kappa 
_{xx}$ is around 24 W m$^{-1}$ K$^{-1}$, which is much higher than those 
of the TE materials for Peiter refrigeration, such as 1.1 W m$^{-1}$ 
K$^{-1}$for Bi$_{2}$Te$_{3}^{34}$, 3.0 W m$^{-1}$ K$^{1\, }$for filled 
skutterudites$^{42}$, and 1.0 W m$^{-1}$ K$^{1}$ for Cu$_{2}$Se$^{39}$. 
However, it is noteworthy that the peak $\kappa_{xx}$ of 
single-crystalline NbSb$_{2}$ is lower than those of many thermomagnetic 
materials for Ettingshausen refrigeration, such as 1290 W m$^{-1}$ K$^{-1\, 
}$for single-crystalline NbP under 8 T$^{43}$, 1586 W m$^{-1}$ K$^{-1\, 
}$for single-crystalline TaP under 9 T$^{20}$, and 215 W m$^{-1}$ K$^{-1\, 
}$for single-crystalline WTe$_{2}$ under 9 T$^{27}$. When the magnetic field 
is applied, the $\kappa_{xx}$ of single-crystalline NbSb$_{2\, }$at low 
temperatures is significantly decreased. As shown in Supplementary Fig. 5, 
the $\kappa_{xx} $at 5 K is 35.9 W m$^{-1}$ K$^{-1}$ when $B =$ 0 T, but 
only 2.7 W m$^{-1}$ K$^{-1}$ when $B \quad =$ 1 T. When the magnetic field is 
increased to 3 T, the $\kappa_{xx} $is further decreased. However, under 
higher magnetic field, the $\kappa_{xx} $is almost unchanged. Such 
$\kappa_{xx}$ reduction under magnetic field is caused by the suppression 
of the contribution of carriers in thermal transports. Moreover, as shown in 
Supplementary Fig. 6, the estimated isothermal $\kappa_{xx}$ is slightly 
smaller than the measured adiabatic $\kappa_{xx}$.

The measured $\kappa_{xx}$ in Fig. 3e is mainly composed of the lattice 
thermal conductivity $\kappa_{l}$ and carrier thermal conductivity 
$\kappa_{e}$. Under magnetic field, their relationship can be expressed 
by the empirical formula$^{20,\, 22,\, 44}$
\begin{equation}
\label{eq2}
\kappa_{xx}\left( {B,T} \right) = \kappa_{l}(T) + \kappa_{e}\left( {B,T} \right) = \kappa_{l}(T) + \frac{\kappa_{e}\left( {0,~~T} \right)}{1 + \eta B^{s}}
\end{equation}
where $\eta $ and $s$ are the two factors related to the thermal mobility and 
scattering mechanism, respectively. The increase of$ B$ will suppress the 
contribution of carriers, which is responsible for the reduction of 
$\kappa_{xx}$ under high magnetic field (Fig. 3e). By using Equation (2), 
the measured $\kappa_{xx}$ data of NbSb$_{2\, }$under different $B$ and $T$ are 
fitted. The fitting results are shown in Supplementary Fig. 5a and 
Supplementary Table 1. The $\kappa_{l}$ increases with increasing 
temperature, reaching the maximum around 25 K, and then decreases at higher 
temperature. The maximum is caused by the transition from the $\kappa 
_{l} $\textasciitilde $T^{3}$ dependence at low temperature to $\kappa 
_{l} $\textasciitilde $T^{-1}$ dependence at high temperature$^{45}$. Based 
on the fitted $\kappa_{e}$, the Lorenz number $L$ can be calculated from the 
Wiedemann-Franz law. As shown in Supplementary Fig. 5b, the $L$ at low 
temperatures is significantly lower than the Sommerfeld value $L =$ 2.44 
\texttimes 1$^{-8}$ W $\Omega $ K$^{-2}$, indicating the violation of 
Wiedemann-Franz law. The ratio of the Lorenz number to Sommerfeld value 
($L$/$L_{0})$ decreases from around 1 near room temperature to the minimum value 
of 0.29 at $T =$ 15 K, and then increases at lower temperature, reaching 0.59 
at 5 K. This trend is similar to the phenomenon found in WP$_{2}$ by Jaoui 
et al.$^{46}$. The violation of WF law might be caused by the inelastic 
scattering of carriers, while the upturn of $L$/$L_{0\, }$below 15 K might be 
caused by the changed carrier scattering mechanism from the inelastic 
scattering into the elastic scattering from the impurities. 

The adiabatic Nernst figure-of-merit $z_{N}$$(=\frac{S_{yx}^{\mathrm{2}}\sigma_{yy}}{\kappa_{xx}}$) under different magnetic fields is shown in 
Fig. 3f. The corresponding adiabatic$ z_{N}T $are shown in Supplementary Fig. 
7a. The $z_{N}$ and $z_{N}T$ increase with increasing temperature, reach the 
peak value around 20 K, and then decrease at higher temperature. Due to the 
enhanced \textit{PF}$_{N}$ and the reduced $\kappa_{xx}$, the $z_{N}$ of 
single-crystalline NbSb$_{2}$ is greatly enhanced by magnetic field. A 
maximum of $z_{N}$ is 33\texttimes 10$^{-4}$ K$^{-1\, }$under 1 T at 15 K, 
which is about six times that of PtSn$_{4}$ under 9 T at 15 K$^{22}$ (Fig. 
1b). The $z_{N}$ is further enhanced to 71\texttimes 10$^{-4}$ K$^{-1}$ under 
5 T at 20 K, corresponding to the adiabatic$ z_{N}T$ of 0.14 and the 
isothermal$ z_{N}T$ of 0.16 (Supplementary Fig. 6b). With further increasing 
the magnetic field, the $z_{N}$ and $z_{N}T$ tend to saturate (Supplementary 
Fig. 7b and Supplementary Fig. 7c). As shown in Fig. 1b, the $z_{N}$ of 
single-crystalline NbSb$_{2}$ is higher than the Peiter figure-of-merit $z$ of 
all the TE materials$^{34,\, 38,\, 40,\, 41,\, 47,\, 48}$. It is among the 
best thermomagnetic materials for Ettingshausen refrigeration reported so 
far. More importantly, the high $z_{N}$ and $z_{N}T$ of NbSb$_{2}$ appear in 
the temperature range of 5-30 K (Fig. 1b and Supplementary Fig. 7d), which 
can well satisfy the requirement of refrigeration below liquid nitrogen 
temperature.

\textbf{Potential application.} Based on the measured thermomagnetic 
properties, the maximum temperature difference ($\Delta T_{max})$ and the 
maximum specific heat pumping power ($P_{max})$ of the present single-crystal 
NbSb$_{2}$ can be estimated by the following equations$^{14,\, 26}$
\begin{equation}
\label{eq3}
\mathrm{\Delta}T_{max} = \frac{1}{2}z_{N}^{iso}T_{c}^{2}
\end{equation}
\begin{equation}
\label{eq4}
P_{max} = \frac{S_{yx}^{2}T_{c}^{2}\sigma_{yy}A}{2lm} = \frac{S_{yx}^{2}T_{c}^{2}\sigma_{yy}}{2Dl^{2}}
\end{equation}
where $z_{N}^{iso}$ is the isothermal figure-of-merit, $T_{c}$ is the 
cold end temperature, $l$ and $A$ are the thickness and cross-sectional area of a 
cuboid sample along the direction of heat flow,$ m$ and $D$ are the mass and 
density of the sample, respectively. Under $B =$ 5 T and $T_{c} =$ 25 
K, the $\Delta T_{max}$ of NbSb$_{2\, }$single crystal is about 2.0 K. 
Particularly, in the condition of B = 5 T and $T_{c}=$ 25 K, the 
theoretical $P_{max}$ of a cuboid sample with $l =$ 1 mm is about 14.2 W 
g$^{-1}$, which is much higher than the compression refrigerator with gas 
refrigerants$^{26}$ (e.g. $P_{max} =$ 0.05 W g$^{-1}$ for He at 5 K, 0.1 W 
g$^{-1}$ for H$_{2}$ at 26 K, and 1.0 W g$^{-1}$ for N$_{2}$ at 93 K). 
Furthermore, the mechanical workability of NbSb$_{2}$ single crystal is very 
good. As shown in Supplementary Fig. 8, it can be easily machined into 
regular thin square and rectangle without cracking. This can facilitate the 
fabrication of the classic exponential shape for Ettingshausen 
refrigeration$^{49}$.

\textbf{Two-carrier model.} The large and unsaturated $S_{yx}$ under high 
magnetic field is indispensable for realizing high \textit{PF}$_{N}$ and $z_{N}$ of 
thermomagnetic materials. As shown in Fig. 4a, beyond the present 
NbSb$_{2}$, nearly all the reported good thermomagnetic materials possess 
such character$^{19,\, 20,\, 22,\, 26,\, 27,\, 50}$. NbSb$_{2\, }$is a 
semimetal with the Fermi level simultaneously crossing the conduction band 
and valence band (Fig. 2c). Thus, both electrons and holes will take part in 
the electrical transports. By using Supplementary Equation (13) and (14), 
the \quad electron (or hole) carrier concentration$ n_{e}$ (or $n_{h})$, and electron 
(or hole) carrier mobility $\mu_{e}$ (or $\mu_{h})$ in NbSb$_{2}$ can 
be obtained by fitting the two-carrier model to the measured transverse 
resistivity $\rho_{xx}(B)$ and Hall resistivity $\rho_{yx}(B)$. This 
model can well fit the $\rho_{xx}(B)$ and $\rho_{yx}(B)$ data over 5-300 
K (Fig. 4b-c). The $n_{e}$ and $n_{h}$ are almost the same with each other 
\textasciitilde 10$^{20}$ cm$^{-3}$ over the entire temperature. Likewise, 
the inset in Fig. 4d shows that the $\mu_{e}$ and $\mu_{h}$ of 
single-crystalline NbSb$_{2}$ are also comparable over the entire 
temperature range. In a two-carrier model$^{51}$ with constant relaxation 
time approximation and under the ideal conditions of $n_{e\, }= n_{h}$ and $\mu_{e}  =  \mu_{h}  =  \overline{\mu}$, the  $S_{yx}$ can be 
expressed as
\begin{equation}
\label{eq5}
S_{yx} = \frac{\overline{\mu}B}{2}\left( S_{xx}^{h} - S_{xx}^{e} \right)
\end{equation}
where $S_{xx}^{e}$ and $S_{xx}^{h}$ are the Seebeck 
thermopower of electrons and holes under the magnetic field $B$, respectively. 
The details about how Equation (5) is obtained can be found in Supplementary 
Note 3. Different from the one-carrier model in which a saturated $S_{yx}$ is 
observed under large magnetic field, the two-carrier model based on Equation 
(5) gives an unsaturated $S_{yx}$ when magnetic field increases, this is 
consistent with the measured $S_{yx}$ \textit{vs.B }behavior of single-crystalline 
NbSb$_{2}$ shown in Fig. 4a.

The inset in Fig. 4d shows that the $\mu_{e}$ and $\mu_{h}$ of 
single-crystalline NbSb$_{2\, }$are very large at low temperature, reaching 
$\mu_{e}=$ 2.1 m$^{2}_{\, }$V$^{-1}$ s$^{-1}$ and $\mu_{h} =$ 
1.2 m$^{2}_{\, }$V$^{-1}$ s$^{-1}$ at 5 K. These values are comparable 
with the high mobility found in the extremely large magnetoresistance 
materials, such as Cd$_{3}$As$_{2}$ ($\mu_{e} =$ 6.5 m$^{2}_{\, 
}$V$^{-1}$ s$^{-1}$ and $\mu_{h} =$ 0.5 m$^{2}_{\, }$V$^{-1}$ 
s$^{-1}$ at 10 K)$^{50}$, PtSn$_{4}$ ($\mu_{e} =$ 7.6 m$^{2}_{\, 
}$V$^{-1}$ s$^{-1}$ and $\mu_{h} =$ 7.6 m$^{2}_{\, }$V$^{-1}$ 
s$^{-1}$ at 2 K)$^{22}$, LaBi ($\mu_{e} =$ 2.6 m$^{2}_{\, }$V$^{-1}$ 
s$^{-1}$ and $\mu_{h} =$ 3.3 m$^{2}_{\, }$V$^{-1}$ s$^{-1}$ at 2 
K)$^{52}$. The observed high $\mu_{e}$ and $\mu_{h}$ are also 
consistent with the Dirac-like band dispersion of NbSb$_{2}$ near the Fermi 
level (Fig. 2c). The large $\mu_{e}$ and $\mu_{h}$ are one important 
reason for the large $S_{yx}$ of single-crystalline NbSb$_{2}$.

In addition, it is instructive to plot $\left( S_{xx}^{h} - S_{xx}^{e} \right)$ of 
single-crystalline NbSb$_{2}$ under different temperatures and magnetic 
fields. In Fig. 4e, $\left( S_{xx}^{h} - S_{xx}^{e} \right)$ shows a local peak at 25 K, which is believed to have a consequence for the observed colossal Nernst power 
factor. In thermoelectrics, such extra-large thermopower peak at low 
temperature is usually caused by the phonon-drag effect$^{14,\, 51}$. With 
increasing temperature, the phonons with higher momentum are excited. When 
the momentum of the long-wave acoustic phonons is similar with that of the 
carriers on the Fermi surface, the phonon-drag effect occurs, leading to the 
appearance of a peak in the Seebeck thermopower curve at low temperature. 
The Seebeck thermopower of a material can be written as $S_{xx} =$ 
$S_{d} +S_{p}$, where $S_{d}$ is related to the charge carrier diffusion 
processes and $S_{p}$ is related to phonons. In a degenerate limit, the 
$S_{d}$ usually has linear temperature dependence$^{53}$. The estimation 
details of $S_{xx}^{e}$ and $S_{xx}^{h}$ are shown in 
Supplementary Note 4. However, as presented in Supplementary Fig. 9, both 
$S_{xx}^{e}$ and $S_{xx}^{h}$ deviate off the linear 
temperature dependence below 100 K, indicating the non-negligible $S_{p}$ in 
single-crystalline NbSb$_{2}$ at low temperatures. By subtracting the 
$S_{d}$ from the $S_{xx}^{e}$ and $S_{xx}^{h}$, the $S_{p}^{e}$ and $S_{p}^{h}$ can be estimated, with the details 
shown in Supplementary Note 5. As shown in Fig. 4f, the absolute values of 
$S_{p}^{e}$ and $S_{p}^{h}$ show the maxima value of 75 $\mu 
$V K$^{-1}$ and 193 $\mu $V K$^{-1}$ around 25 K, much larger than the$S_{d}^{e}$(5.4 $\mu $V K$^{-1})$ and $S_{d}^{h}$ (3.6 $\mu $V 
K$^{-1})$ at the same temperature, respectively. Consequently, the 
synergistic effect of $S_{p}^{e}$ and $S_{p}^{h}$ greatly 
improves the total Ettingshausen effect in the single-crystalline 
NbSb$_{2}$. At higher temperature, the phonon-drag effect is quickly 
weakened since the significantly excited high-frequency phonons lead to the 
reduction of the relaxation time of long-wave acoustic phonons$^{14}$. Thus, 
the $S_{p}^{e}$ and $S_{p}^{h}$ are quickly decreased after 
reaching the maxima values. Above 125 K, the electrical transports are 
mainly determined by the carrier diffusion process. At this time, the 
measured $S_{yx}$ is comparable with the theoretical value of $283\overline{\mu}/E_{F}T$ (Supplementary Fig. 10)$^{54}$, where $E_{F} =$ 1200 K is 
derived from the relation $E_{F} = \frac{\hbar^{2}}{2m}\left( 3\pi^{2}n \right)^{2/3}$ $^{\, 55}$, with the carrier 
concentration $n$ equaling to 1.5\texttimes 10$^{20}$ cm$^{-3}$ and $m$ equaling 
to the free electron mass $m_{0}$. These prove that the fitted $\mu_{\, 
}$and $n$ in Fig. 4d are reasonable. \\

\section*{Discussion}

In summary, we report a colossal Nernst power factor of 3800\texttimes 
10$^{-4}$ W m$^{-1}$ K$^{-2}$ under 5 T at 25 K and a high Nernst 
figure-of-merit $z_{N}$ with of 71\texttimes 10$^{-4}$ K$^{-1}$ under 5 T at 
20 K in single-crystalline NbSb$_{2}$. There are a number of factors 
synergistically contributed to the large and unsaturated Nernst thermopower 
$S_{yx}$ under magnetic field: 1) a favorable band structure providing nearly 
identical electron and hole concentrations at Fermi level, 2) extraordinary 
high electron and hole mobilities benefiting from the Dirac-like dispersion 
of low energy excitations common to several well-known topological 
semimetals, and 3) strong phonon-drag effect. The phonon-drag effect derived 
from our data analysis suggests phonon can play an important role in the 
transport process of Dirac fermions, which is another interesting phenomenon 
worth of further investigation. This work provides a new material option for 
the solid-state heat pumping below liquid nitrogen temperature. 

\section*{Methods}

\textbf{Sample synthesis.} NbSb$_{2}$ single crystal was synthesized by 
chemical vapor transport method in two steps. First, polycrystalline powder 
was synthesized by solid-stated reaction. The niobium powder (alfa, 
99.99{\%}) and antimony shot (alfa, 99.9999{\%}) with stoichiometry 1:2 was 
encapsulated in a vacuum quartz tube and reacted at 1023 K for 48 h. Next, 
the polycrystalline NbSb$_{2}$ powders and 0.3 g iodine were sealed in 
another vacuum quartz tube. The quartz tube was placed in a horizontal 
furnace with a temperature gradient for 2 weeks. The hot end temperature and 
cold end temperature of the quartz tube are 1373 K and 1273 K, respectively. 
Finally, shiny and bar-like single crystals appear in the cold end of the 
quartz tube. 

\textbf{Characterization and transport property measurements.} The phase 
composition of the single-crystalline NbSb$_{2}$ was characterized by X-ray 
diffraction (XRD, D/max-2550 V, Rigaku, Japan) and scanning electron 
microscopy (SEM, ZEISS supra-55, Germany) with an energy dispersive X-ray 
spectroscopy (EDS, Oxford, UK). The electrical and thermal transport 
properties of single-crystalline NbSb$_{2}$ were measured under the magnetic 
field by using physical property measurement system (PPMS, Quantum design, 
USA). The alternating current was used in the electrical conductivity 
measurement with the purpose to eliminate the thermal Hall effect. The 
transverse resistivity and Hall resistivity were measured by four-probe 
method and five-probe method, respectively. The Seebeck thermopower was 
measured on a standard thermal transport option (TTO) platform. The Nernst 
thermopower was measured on a modified TTO platform, where the Cu wires for 
measuring voltage signals were separated from the Cernox 1050 thermometers. 
All measurements of thermal transport were performed by using four-probe 
method. The details can be found in the Supplementary Note 6 and 
Supplementary Figs. 11a-b. The measurement direction was marked in the inset 
of Fig. 2b, which was the same as that of the Seebeck thermopower. Taking 
$b$-axis as the $x$ direction and $c$-axis as the $y$ direction, the magnetic field was 
applied in the $z$ direction perpendicular to the \textit{bc} plane. In addition, via 
comparing with the thermal conductivity of the sample with and without 
adhering Cu wires (Supplementary Fig. 12), it is concluded that the Cu wires 
have little influence on the measurement.

\textbf{Calculation.} First-principles calculations were carried out using 
Quantum espresso software package$^{56}$ with the lattice parameters given 
in the materials project$^{57}$. Perdew-Burke-Ernzerhof (PBE) 
exchange-correlation functional$^{58}$ within the generalized gradient 
approximation (GGA) and fully relativistic norm-conserving pseudopotentials 
generated using the optimized norm-conserving Vanderbilt 
pseudopotentials$^{59}$ were used in the calculations. The primitive 
Brillouin zone was sampled by using a 10 \texttimes 10 \texttimes 10 
Monkhorst-Pack $k$ mesh and a plane-wave energy cut off of 900 eV was used. The 
Fermi surface calculation was performed on a dense $k$ mesh of 41 \texttimes 41 
\texttimes 41 and was visualized by using XCrysDen software$^{60}$. The QE 
calculations were also verified using the projector-augmented wave 
(PAW)$^{61}$ method as implemented in the Vienna ab initio simulation 
package (VASP)$^{62}$ which gave similar results.

\subsection*{Data availability}

The data generated in this study are provided in the Source Data file.\\

\noindent \textbf{References}\\

1. Yan, Q. {\&} Kanatzidis, M. G. High-performance thermoelectrics and 
challenges for practical devices. \textit{Nat. Mater.} \textbf{21}, 503-513 (2022).

2. Pei, Y., Shi, X., LaLonde, A., Wang, H., Chen, L. {\&} Snyder, G. J. 
Convergence of electronic bands for high performance bulk thermoelectrics. 
\textit{Nature} \textbf{473}, 66-69 (2011).

3. Fu, C., Zhu, T., Liu, Y., Xie, H. {\&} Zhao, X. Band engineering of high 
performance p-type FeNbSb based half-Heusler thermoelectric materials for 
figure of merit zT \textgreater 1. \textit{Energy Environ. Sci.} \textbf{8}, 216-220 (2015).

4. Mao, J., Chen, G. {\&} Ren, Z. Thermoelectric cooling materials. \textit{Nat. Mater.} 
\textbf{20}, 454-461 (2021).

5. Jiang, B., et al. High-entropy-stabilized chalcogenides with high 
thermoelectric performance. \textit{Science} \textbf{371}, 830-834 (2021).

6. Gr\"{o}blacher, S., et al. Demonstration of an ultracold 
micro-optomechanical oscillator in a cryogenic cavity. \textit{Nat. Phys.} \textbf{5}, 485-488 
(2009).

7. Hornibrook, J. M., et al. Cryogenic Control Architecture for Large-Scale 
Quantum Computing. \textit{Phys. Rev. Appl.} \textbf{3}, 024010 (2015).

8. Collaudin, B. {\&} Rando, N. Cryogenics in space: a review of the 
missions and of the technologies. \textit{Cryogenics} \textbf{40}, 797-819 (2000).

9. Zhang, Q., Bai, S. {\&} Chen, L. Technologies and Applications of 
Thermoelectric Devices: Current Status, Challenges and Prospects. \textit{J. Inorg. Mater.} 
\textbf{34}, 279-293 (2018).

10. Chu, J., et al. Electrode interface optimization advances conversion 
efficiency and stability of thermoelectric devices. \textit{Nat. Commun.} \textbf{11}, 2723 
(2020).

11. Qiu, P., et al. High-Efficiency and Stable Thermoelectric Module Based 
on Liquid-Like Materials. \textit{Joule} \textbf{3}, 1538-1548 (2019).

12. Xing, T., et al. High efficiency GeTe-based materials and modules for 
thermoelectric power generation. \textit{Energy Environ. Sci.} \textbf{14}, 995-1003 (2021).

13. Mahan, G. D. Figure of merit for thermoelectrics. \textit{J. Appl. Phys.} \textbf{65}, 
1578-1583 (1989).

14. Goldsmid, H. J. \textit{Thermoelectric Refrigeration}. (Plenum Press, Wembley, 1964).

15. v. Ettingshausen, A. {\&} Nernst, W. Ueber das Auftreten 
electromotorischer Kr\"{a}fte in Metallplatten, welche von einem 
W\"{a}rmestrome durchflossen werden und sich im magnetischen Felde befinden. 
\textit{Ann. Phys.} \textbf{265}, 343-347 (1886).

16. Cuff, K. F., Horst, R. B., Weaver, J. L., Hawkins, S. R., Kooi, C. F. 
{\&} Enslow, G. M. The thermomagnetic figure of merit and Ettingshausen 
cooling in Bi--Sb alloys. \textit{Appl. Phys. Lett.} \textbf{2}, 145-146 (1963).

17. Yim, W. M. {\&} Amith, A. Bi-Sb alloys for magneto-thermoelectric and 
thermomagnetic cooling. \textit{Solid-State Electron.} \textbf{15}, 1141-1165 (1972).

18. Madon, B., Wegrowe, J. E., Drouhin, H. J., Liu, X., Furdyna, J. {\&} 
Khodaparast, G. A. Influence of the carrier mobility distribution on the 
Hall and the Nernst effect measurements in n-type InSb. \textit{J. Appl. Phys.} \textbf{119}, 
025701 (2016).

19. Watzman, S. J., et al. Dirac dispersion generates unusually large Nernst 
effect in Weyl semimetals. \textit{Phys. Rev. B} \textbf{97}, 161404(R) (2018).

20. Han, F., et al. Quantized thermoelectric Hall effect induces giant power 
factor in a topological semimetal. \textit{Nat. Commun.} \textbf{11}, 6167 (2020).

21. Wang, P., et al. Giant Nernst effect and field-enhanced transversal 
$z_{N}T$ in ZrTe$_{5}$. \textit{Phys. Rev. B} \textbf{103}, 045203 (2021).

22. Fu, C., et al. Largely suppressed magneto-thermal conductivity and 
enhanced magneto-thermoelectric properties in PtSn$_{4}$. \textit{Research} \textbf{2020}, 
4643507 (2020).

23. Fu, C., et al. Large Nernst power factor over a broad temperature range 
in polycrystalline Weyl semimetal NbP. \textit{Energy Environ. Sci.} \textbf{11}, 2813-2820 (2018).

24. Liu, W., et al. Weyl Semimetal States Generated Extraordinary 
Quasi-Linear Magnetoresistance and Nernst Thermoelectric Power Factor in 
Polycrystalline NbP. \textit{Adv. Funct. Mater.} \textbf{32}, 2202143 (2022).

25. Feng, T., et al. Large Transverse and Longitudinal 
Magneto-Thermoelectric Effect in Polycrystalline Nodal-Line Semimetal 
Mg$_{3}$Bi$_{2}$. \textit{Adv. Mater.} \textbf{34}, 2200931 (2022).

26. Chen, Z., et al. Leveraging bipolar effect to enhance transverse 
thermoelectricity in semimetal Mg$_{2}$Pb for cryogenic heat pumping. \textit{Nat. Commun.} 
\textbf{12}, 3837 (2021).

27. Pan, Y., He, B., Helm, T., Chen, D., Schnelle, W. {\&} Felser, C. 
Ultrahigh transverse thermoelectric power factor in flexible Weyl semimetal 
WTe$_{2}$. \textit{Nat. Commun.} \textbf{13}, 3909 (2022).

28. Li, Q., et al. Chiral magnetic effect in ZrTe$_{5}$. \textit{Nat. Phys.} \textbf{12}, 
550-554 (2016).

29. Skinner, B. {\&} Fu, L. Large, nonsaturating thermopower in a quantizing 
magnetic field. \textit{Sci. Adv.} \textbf{4}, eaat2621 (2018).

30. Kang, J. S., Vu, D. {\&} Heremans, J. P. Identifying the Dirac point 
composition in Bi$_{1-x}$Sb$_{x}$ alloys using the temperature dependence of 
quantum oscillations. \textit{J. Appl. Phys.} \textbf{130}, 225106 (2021).

31. Lee, S. E., et al. Orbit topology analyzed from $\pi $ phase shift of 
magnetic quantum oscillations in three-dimensional Dirac semimetal. \textit{Proc. Natl. Acad. Sci. U.S.A.} 
\textbf{118}, e2023027118 (2021).

32. Furuseth, S. {\&} Kjekshus, A. Arsenides and Antimonides of Niobium. 
\textit{Nature} \textbf{203}, 512-512 (1964).

33. Wang, K., Graf, D., Li, L., Wang, L. {\&} Petrovic, C. Anisotropic giant 
magnetoresistance in NbSb$_{2}$. \textit{Sci. Rep.} \textbf{4}, 7328 (2014).

34. Mao, J., et al. High thermoelectric cooling performance of n-type 
Mg$_{3}$Bi$_{2}$-based materials. \textit{Science} \textbf{365}, 495-498 (2019).

35. Ali, M. N., et al. Large, non-saturating magnetoresistance in WTe$_{2}$. 
\textit{Nature} \textbf{514}, 205-208 (2014).

36. Shekhar, C., et al. Extremely large magnetoresistance and ultrahigh 
mobility in the topological Weyl semimetal candidate NbP. \textit{Nat. Phys.} \textbf{11}, 
645-649 (2015).

37. Mun, E., Ko, H., Miller, G. J., Samolyuk, G. D., Bud'ko, S. L. {\&} 
Canfield, P. C. Magnetic field effects on transport properties of 
PtSn$_{4}$. \textit{Phys. Rev. B} \textbf{85}, 035135 (2012).

38. Liang, J., et al. Crystalline Structure-Dependent Mechanical and 
Thermoelectric Performance in Ag$_{2}$Se$_{1-x}$S$_{x}$ System. \textit{Research} 
\textbf{2020}, 6591981 (2020).

39. Liu, H., et al. Copper ion liquid-like thermoelectrics. \textit{Nat. Mater.} \textbf{11}, 
422-425 (2012).

40. Wang, L., et al. Discovery of low-temperature GeTe-based thermoelectric 
alloys with high performance competing with Bi$_{2}$Te$_{3}$. \textit{J. Mater. Chem. A} \textbf{8}, 
1660-1667 (2020).

41. Qin, B., et al. Power generation and thermoelectric cooling enabled by 
momentum and energy multiband alignments. \textit{Science} \textbf{373}, 556-561 (2021).

42. Shi, X., et al. Multiple-Filled Skutterudites: High Thermoelectric 
Figure of Merit through Separately Optimizing Electrical and Thermal 
Transports. \textit{J. Am. Chem. Soc.} \textbf{133}, 7837-7846 (2011).

43. Stockert, U., et al. Thermopower and thermal conductivity in the Weyl 
semimetal NbP. \textit{J. Phys.: Condens. Matter} \textbf{29,} 325701 (2017).

44. Oca\~{n}a, R. {\&} Esquinazi, P. Thermal conductivity tensor in 
YBa$_{2}$Cu$_{3}$O$_{7-x}$ : Effects of a planar magnetic field. \textit{Phys. Rev. B} 
\textbf{66}, 064525 (2002).

45. Ashcroft, N. W. {\&} David, M. N. \textit{Solid state physics}. (Saunders College Publishing, New 
York, 1976).

46. Jaoui, A., et al. Departure from the Wiedemann--Franz law in WP$_{2}$ 
driven by mismatch in T-square resistivity prefactors. \textit{npj Quantum Mater.} \textbf{3}, 64 
(2018).

47. Lenoir, B., Cassart, M., Michenaud, J. P., Scherrer, H. {\&} Scherrer, 
S. Transport properties of Bi-RICH Bi-Sb alloys. \textit{J. Phys. Chem. Solids} \textbf{57}, 89-99 (1996).

48. Xu, Q., et al. Thermoelectric properties of phosphorus-doped van der 
Waals crystal Ta$_{4}$SiTe$_{4}$. \textit{Mater. Today Phys.} \textbf{19}, 100417 (2021).

49. Scholz, K., Jandl, P., Birkholz, U. {\&} Dashevskii, Z. M. Infinite 
stage Ettingshausen cooling in Bi-Sb alloys. \textit{J. Appl. Phys.} \textbf{75}, 5406-5408 (1994).

50. Xiang, J., et al. Large transverse thermoelectric figure of merit in a 
topological Dirac semimetal. \textit{Sci. China: Phys., Mech. Astron.} \textbf{63}, 237011 (2019).

51. Delves, R. T. Figure of merit for Ettingshausen cooling. \textit{Br. J. Appl. Phys.} \textbf{15}, 
105-106 (1964).

52. Sun, S., Wang, Q., Guo, P., Liu, K. {\&} Lei, H. Large magnetoresistance 
in LaBi: origin of field-induced resistivity upturn and plateau in 
compensated semimetals. \textit{New J. Phys.} \textbf{18}, 082002 (2016).

53. Blatt, F. J., Schroeder, P. A., Foiles, C. L. {\&} Greig, D. 
\textit{Thermoelectric power of metals}. (Plenum Press, New York and London, 1976).

54. Behnia, K. {\&} Aubin, H. Nernst effect in metals and superconductors: a 
review of concepts and experiments. \textit{Rep. Prog. Phys.} \textbf{79}, 046502 (2016).

55. Gould, H. {\&} Tobochnik, J. \textit{Statistical and Thermal Physics with Computer Application}. (Priceton University Press, Priceton, 
2010).

56. Giannozzi, P., et al. QUANTUM ESPRESSO: a modular and open-source 
software project for quantum simulations of materials. \textit{J. Phys.: Condens. Matter} \textbf{21}, 395502 
(2009).

57. Jain, A., et al. Commentary: The Materials Project: A materials genome 
approach to accelerating materials innovation. \textit{APL Mater.} \textbf{1}, 011002 (2013).

58. Perdew, J. P., Burke, K. {\&} Ernzerhof, M. Generalized Gradient 
Approximation Made Simple. \textit{Phys. Rev. Lett.} \textbf{77}, 3865-3868 (1996).

59. Hamann, D. R. Optimized norm-conserving Vanderbilt pseudopotentials. 
\textit{Phys. Rev. B} \textbf{88}, 085117 (2013).

60. Kokalj, A. XCrySDen---a new program for displaying crystalline 
structures and electron densities. \textit{J. Mol. Graphics Modell.} \textbf{17}, 176-179 (1999).

61. Bl\"{o}chl, P. E. Projector augmented-wave method. \textit{Phys. Rev. B} \textbf{50}, 
17953-17979 (1994).

62. Kresse, G. {\&} Hafner, J. Ab initio molecular dynamics for liquid 
metals. \textit{Phys. Rev. B} \textbf{47}, 558-561 (1993).\\

\noindent \textbf{Acknowledgements }

This work is supported by the National Natural Science Foundation of China 
(91963208, L.C. and 52122213, P.Q.), and Shanghai Government (20JC1415100, 
X.S.). This work at Brookhaven National Laboratory was supported by U.S. 
Department of Energy (DOE) the Office of Basic Energy Sciences, Materials 
Sciences and Engineering Division under Contract No. DE-SC0012704.\\

\noindent \textbf{Author contributions} 

P.L., P.Q., and X.S. designed the experiment. P.L. synthesized the samples 
and performed the transport property measurements, with the help of Q.X. and 
J.X., and N.A. provided band structure calculations. P.L., Q.X., J.L., and 
Y. X. analyzed the transport properties based on the two-carrier model. 
P.L., P.Q., Q.L., L.C., and X.S. analyzed the data and wrote the manuscript.\\

\noindent \textbf{Competing interests}

The authors declare no competing interests.\\
\newpage
\noindent \textbf{Figures}\\

\graphicspath{ {./images/} }

\begin{figure}[h]
\includegraphics[width=1\textwidth]{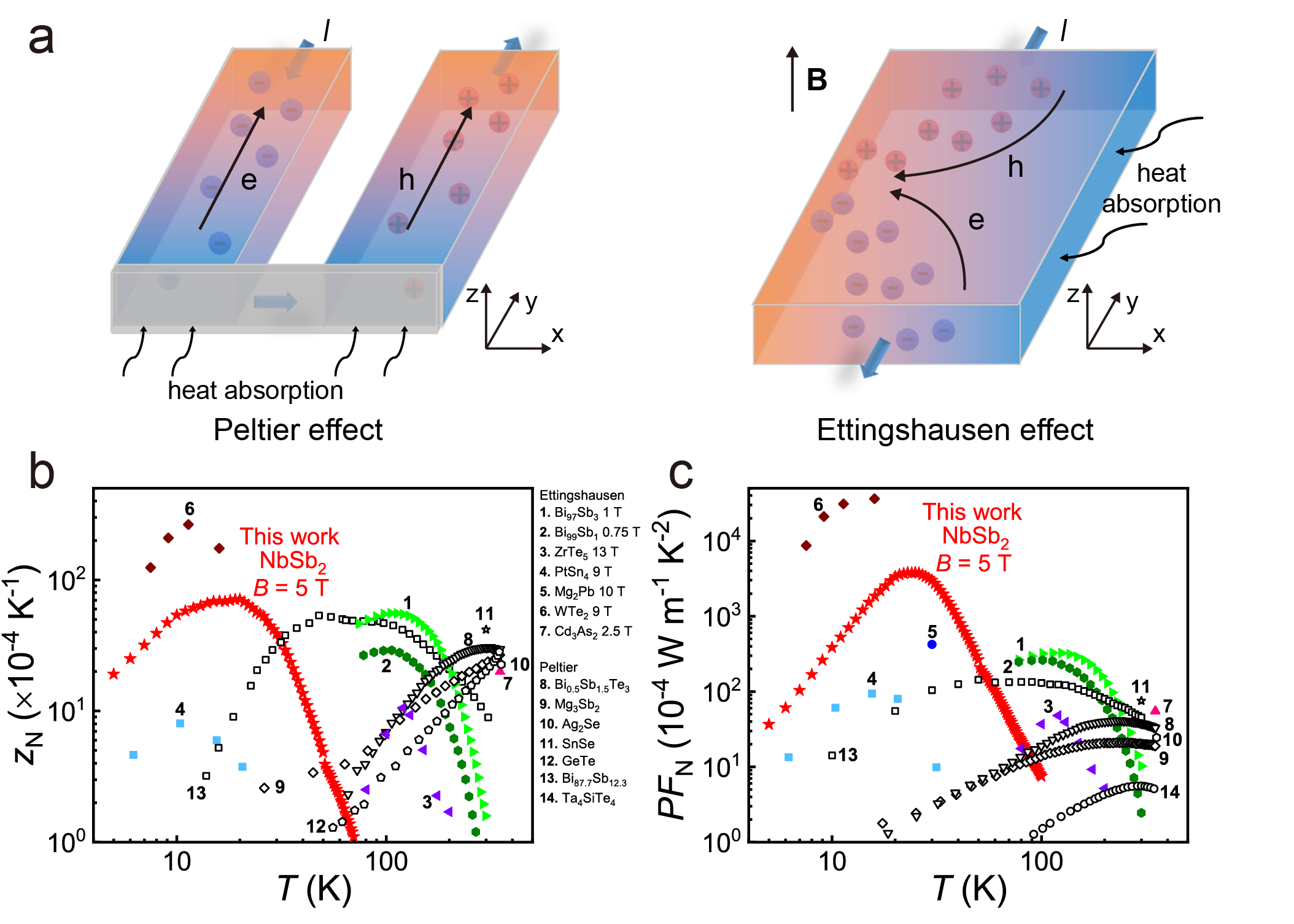}
Figure 1.\textbf{ Peltier effect and Ettingshausen effect.} \textbf{a} 
Schematic diagrams of the Peltier effect and Ettingshausen effect. 
Comparison of \textbf{b} the Nernst figure-of-merit ($z_{N})$ and \textbf{c} 
Nernst power factor (\textit{PF}$_{N})$ for single-crystalline NbSb$_{2}$ and other 
thermomagnetic materials$^{16,\, 17,\, 21,\, 22,\, 26,\, 27,\, 50}$. The 
Peltier figure-of-merit ($z)$ and Peltier power factor (\textit{PF}) of typical TE 
materials are also included$^{34,\, 38,\, 40,\, 41,\, 47,\, 48}$.
\end{figure}
\graphicspath{ {./images/} }
\begin{figure}
\includegraphics[width=1\textwidth]{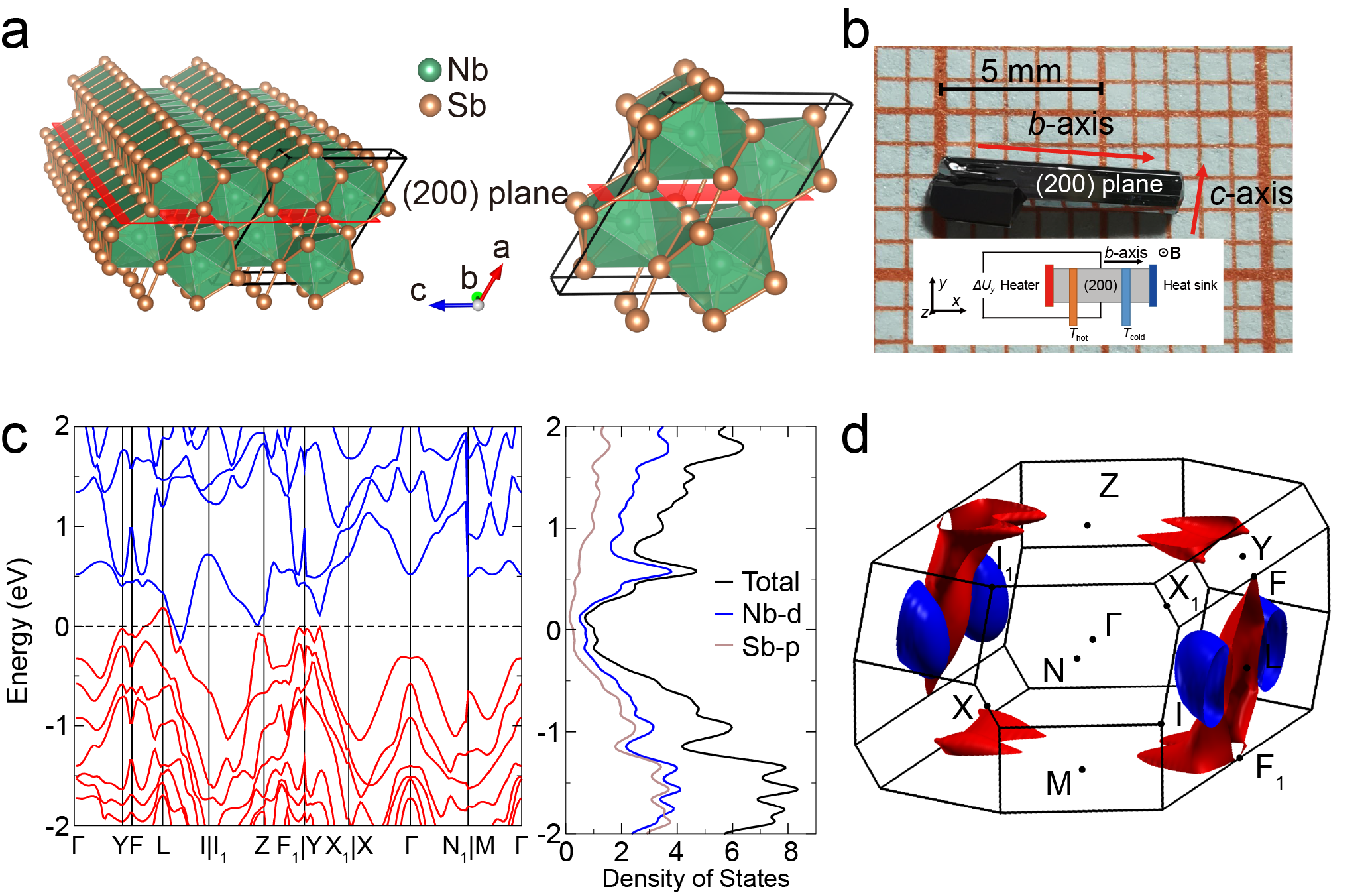}

Figure 2.\textbf{ Crystal structure and band structure of 
NbSb}$_{\mathbf{2}}$\textbf{. a} Crystal structure of NbSb$_{2}$ from 
different perspectives. \textbf{b} Optical image of NbSb$_{2}$ single 
crystal grown in this work. The inset shows the measurement direction of 
Nernst thermopower. \textbf{c} Calculated band structure, density of states, 
and \textbf{d} Fermi surface with the spin-orbit coupling (SOC) for 
NbSb$_{2}$. The red and blue pockets denote the hole and electron pockets, 
respectively.
\end{figure}
\graphicspath{ {./images/} }
\begin{figure}
\includegraphics[width=1\textwidth]{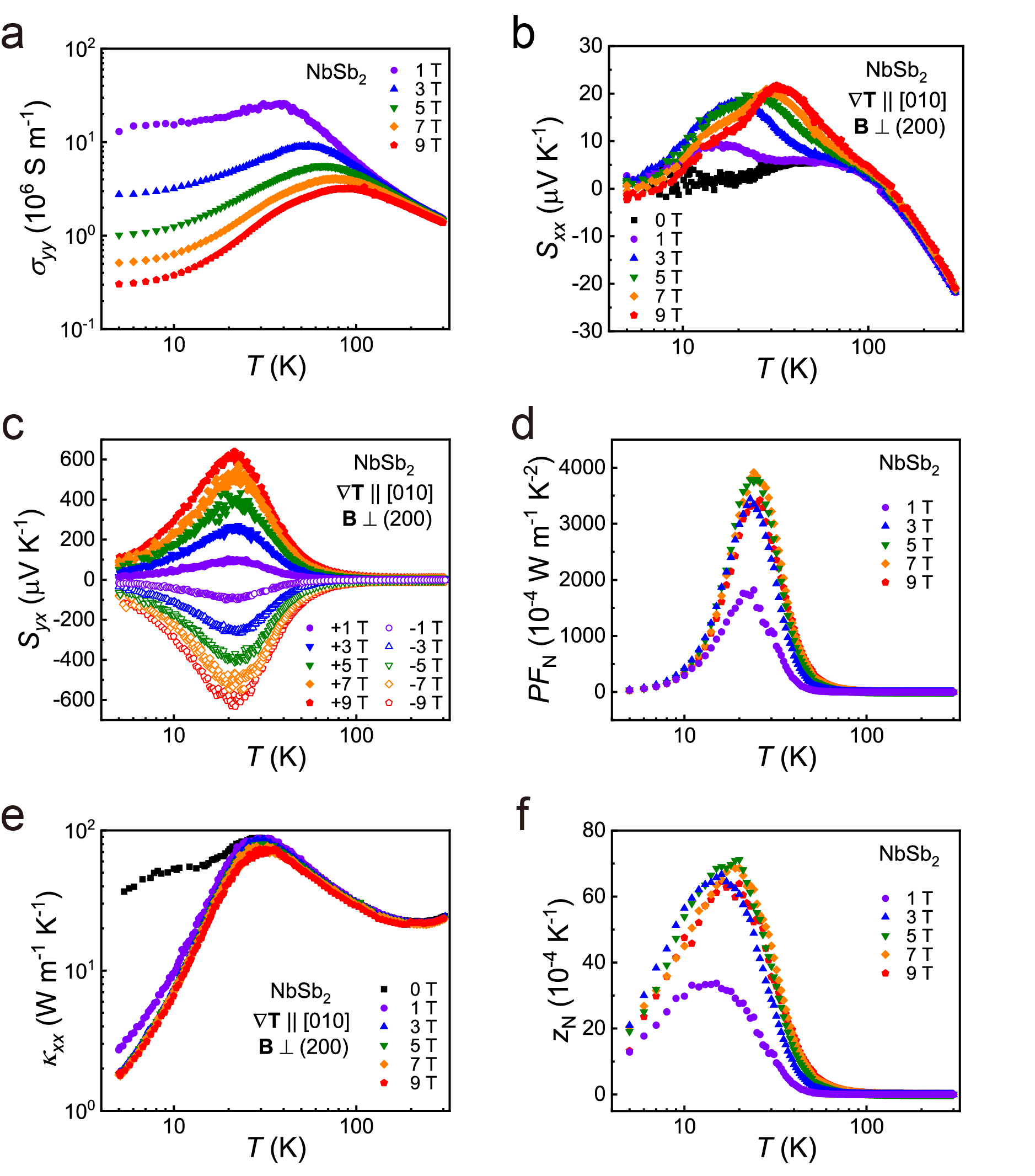}

Figure 3.\textbf{ Electrical and thermal properties of 
NbSb}$_{\mathbf{2}}$\textbf{ for Ettingshausen refrigeration.} Temperature 
dependences of \textbf{a} electrical conductivity ($\sigma_{yy})$, 
\textbf{b} Seebeck thermopower ($S_{xx})$, \textbf{c} Nernst thermopower 
($S_{yx})$, \textbf{d} Nernst power factor (\textit{PF}$_{N})$, \textbf{e} transverse 
thermal conductivity ($\kappa_{xx})$, and \textbf{f} Nernst 
figure-of-merit ($z_{N})$ of single-crystalline NbSb$_{2}$ under different 
magnetic fields and adiabatic condition. In thermal transport measurements, 
the temperature gradient $\nabla$\textbf{T} is parallel to the 
[010] direction and the magnetic field \textbf{B} is perpendicular to the 
(200) plane.
\end{figure}
\graphicspath{ {./images/} }
\begin{figure}
\includegraphics[width=0.9\textwidth]{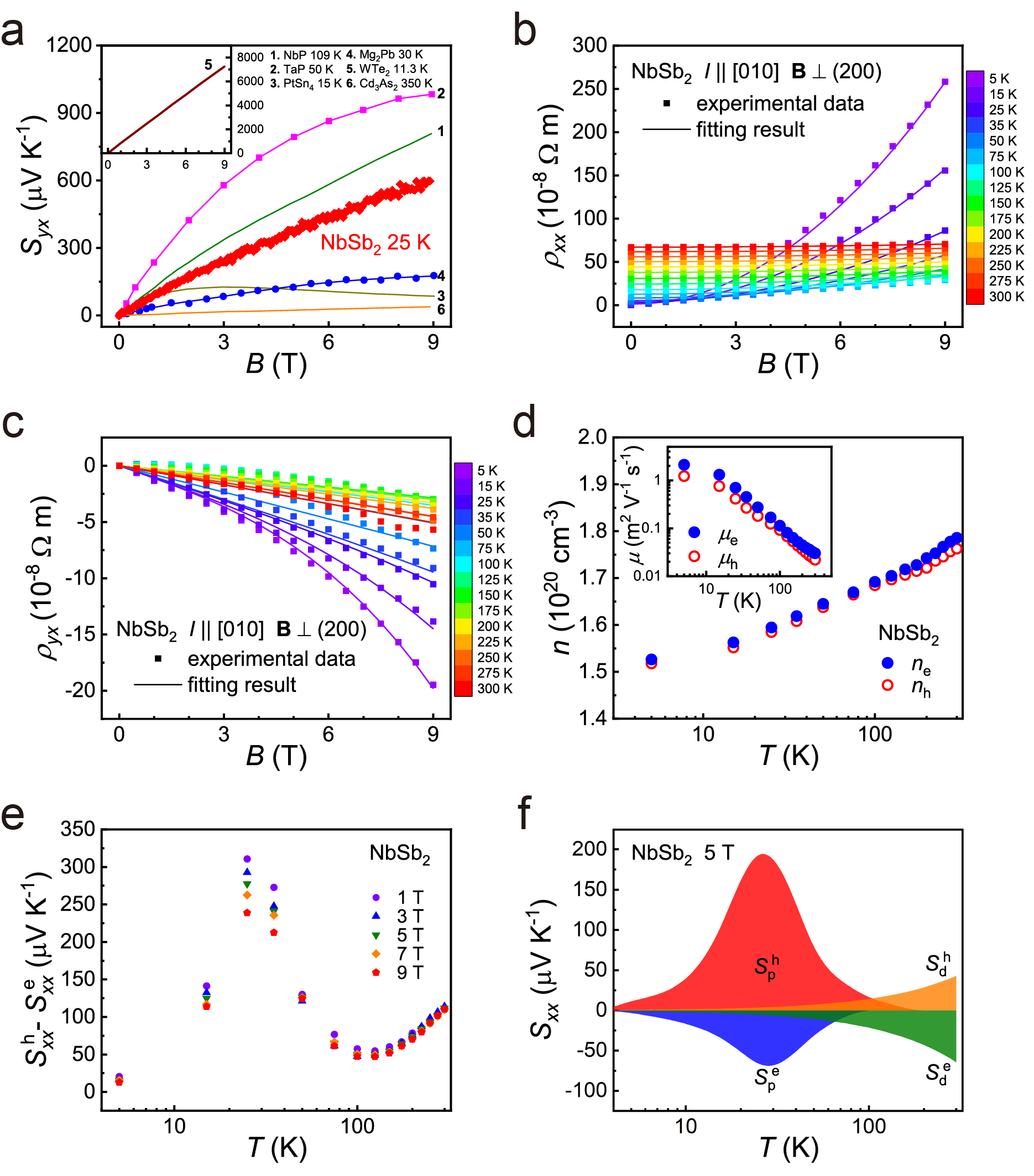}

Figure 4.\textbf{ Detailed electrical transports for 
NbSb}$_{\mathbf{2}}$\textbf{ single crystal.} \textbf{a} Nernst thermopower 
($S_{yx})$ of single-crystalline NbSb$_{2}$ as a function of magnetic field 
$B$ at 25 K. The data for PtSn$_{4}^{22}$, Cd$_{3}$As$_{2}^{50}$, 
Mg$_{2}$Pb$^{26}$, NbP$^{19}$, TaP$^{20}$, and WTe$_{2}^{27}$ are included 
for comparison. \textbf{b} Fitting of the transverse resistivity $\rho 
_{xx}(B)$ and \textbf{c} Hall resistivity $\rho_{yx}(B)$ of 
single-crystalline NbSb$_{2}$ under different temperatures. The symbols are 
experimental data and the lines are the fitting curves. In electrical 
transport measurements, the current $I$ is parallel to the [010] direction and 
the magnetic field \textbf{B} is perpendicular to the (200) plane. 
\textbf{d} Carrier concentrations ($n_{e}$ and $n_{h})$ and carrier mobilities 
($\mu_{e}$ and $\mu_{h})$ of single-crystalline NbSb$_{2}$. \textbf{e} 
Temperature dependence of the difference between Seebeck thermopower of 
electrons and holes $( S_{xx}^{h} - S_{xx}^{e})$ of single-crystalline NbSb$_{2}$ under different magnetic fields. \textbf{f }Seebeck thermopower of electrons 
and holes related to the charge carrier diffusion processes ($S_{p}^{e}$ and $S_{p}^{h}$) and phonons ($S_{d}^{e}$ and $S_{d}^{h}$) at 5 T, respectively. 
\end{figure}

\end{document}